\documentstyle[12pt,epsf]{article}
\newcommand{\barr}{\begin{array}}
\newcommand{\earr}{\end{array}}
\newcommand{\beq}{\begin{equation}}
\newcommand{\eeq}{\end{equation}}
\newcommand{\bea}{\begin{eqnarray}}
\newcommand{\eea}{\end{eqnarray}}

\newcommand{\cmpr}[3]{Commun.\ Math.\ Phys.\ {\bf #1}, #2 (#3)}
\newcommand{\prlr}[3]{Phys.\ Rev.\ Lett.\ {\bf #1}, #2 (#3)}

\newcommand{\prbr}[3]{Phys.\ Rev.\ B~{\bf #1}, #2 (#3)}
\newcommand{\prdr}[3]{Phys.\ Rev.\ D~{\bf #1}, #2 (#3)}
\newcommand{\plbr}[3]{Phys.\ Lett.\ {\bf #1}, #2 (#3)}
\newcommand{\npbr}[3]{Nucl.\ Phys.\ {\bf #1}, #2 (#3)}
\newcommand{\rmpr}[3]{Rev.\ Mod.\ Phys. {\bf #1}, #2 (#3)}
\newcommand{\jpcr}[3]{J.\ Phys.\ {\bf #1}, #2 (#3)}

\makeatletter

\def\compoundrel#1\over#2{\mathpalette\compoundreL{{#1}\over{#2}}}
\def\compoundreL#1#2{\compoundREL#1#2}
\def\compoundREL#1#2\over#3{\mathrel
	{\vcenter{\hbox{$\m@th\buildrel{#1#2}\over{#1#3}$}}}}
\makeatother

\def\c+{c^{\dagger}}
\def\d+{d^{\dagger}}

\include{psfig}

\begin{document}

\title{Quantum Critical Behavior in Gauged Yukawa Matrix Field Theories with 
Quenched Disorder}
\author{H. Hamidian}

\maketitle
\begin{center}
{\sl Department of Physics, Stockholm University, Box 6730, S-113~85 Stockholm, 
Sweden}
\end{center}

\begin{abstract}

\thispagestyle{empty}

We use the Wilson-Fisher $\epsilon$ expansion to study quantum critical behavior 
in gauged Yukawa matrix field theories with weak quenched disorder. We find that 
the resulting quantum critical behavior is in the universality class of the pure 
system. As in the pure system, the phase transition is typically first order, 
except for a limited range of parameters where it can be second order with 
computable critical exponents. Our results apply to the study of two-dimensional 
quantum antiferromagnets with weak quenched disorder and provide an example for 
fluctuation-induced first order phase transitions in circumstances where naively 
none is expected.    
\end{abstract}

\newpage

One of the most remarkable features of quantum spin systems is their 
relationship to gauge theories. This connection, which was originally used to 
study chiral symmetry breaking in QCD in the strong coupling limit 
\cite{smit}, remains a powerful tool.
Recently this analogy has been used to prove that certain gauge theories break 
chiral symmetry in the strong coupling limit
[2-4]. 
Also, it has been further suggested 
\cite{beran} 
that the underlying analogy can be used in a much broader sense to account for 
the quasi-particle spectrum and other infrared features of two-dimensional (2D) 
quantum antiferromagnets and three-dimensional non-Abelian gauge theories. Using 
the quantum chromodynamics (QCD) terminology, the quantum phase transition with 
a chiral symmetry breaking pattern is then characterized by the number of 
flavors ($N_F$) and colors ($N_C$) of quarks in the gauge theory. In the 
antiferromagnet the rank of the spin algebra and the size of its representation 
play the same role as the number of flavors and colors respectively.

The study of 2D quantum antiferromagnets, particulary in connection with 
high-$T_C$ superconductivity, is a rapidly developing subject of great current 
interest 
\cite{barzykin}. 
An important question that naturally arises is the stability of the 
antiferromagnetic long-range order (LRO) in the presence of quenched disorder 
(QD). In fact, since the N\'{e}el temperature of planar high-$T_C$ 
superconductors, such as ${\rm La_2CuO_4}$, is extremely sensitive to impurities 
and defects, it is probably not correct to ignore the impurity effects even in 
high-quality samples used in laboratories. The critical behavior of 
low-dimensional systems with quenched disorder and an $O(N)$-vector magnetic 
order parameter has been addressed by a number of authors
\cite{dorogovtsev,boyanovsky} 
by using a Landau-Ginzburg-Wilson (LGW) Hamiltonian and applying the 
Wilson-Fisher (WF) $\epsilon$ expansion 
\cite{wilson}. In particular, Boyanovsky and Cardy \cite{boyanovsky} have 
carried out a two-loop (double) $\epsilon$ expansion in systems in which the 
impurities are correlated in $\epsilon_d$ dimensions and randomly distributed in 
$d-\epsilon_d$ dimensions. They take the full structure of the theory into 
account which, as a result of anisotropies, leads to highly nonlocal 
interactions. 

In this paper we shall study the stability of LRO in 2D quantum antiferromagnets 
with QD by using the WF $\epsilon$ expansion and renormalization group (RG) 
techniques. We shall approach this problem by examining the critical behavior of 
the related zero temperature gauged Yukawa matrix field theories in three 
(Euclidean) dimensions in the presence of quenched disorder. This requires that 
the anisotropies be properly taken into account, as is done (see Ref. 
\cite{boyanovsky}, {\em e.g.}) for the $O(N)$-vector model in the presence of 
QD. However we shall go one step further and also examine the stability of the 
theory against quantum fluctuations which can arise through the Coleman-Weinberg 
mechanism \cite{coleman}. Our results enable us to argue that below a certain 
critical doping (impurity concentration) $\Delta_{\rm crit.}$, and in a narrow 
region in parameter space, 2D doped quantum antiferromagnets belong to the 
universality class of gauged Yukawa matrix field theories which describe the 
pure system. We check this result through an explicit calculation of the 
critical exponent $\nu$ and find that the Harris criterion 
\cite{harris} 
is indeed satisfied. However, we find that a careful examination of the RG flows 
in the space of couplings indicates that LRO can be destroyed due to quantum 
fluctuations. This means that there are cases where there is an infrared (IR) 
stable fixed point and the Harris criterion is satisfied, but the phase 
transition is fluctuation-induced {\it first order} rather than {\it second 
order}, contrary to what is naively expected. We shall begin with a brief review 
of some recent work on pure quantum antiferromagnets and gauged Yukawa matrix 
theories.

The pure 2D (generalized) quantum antiferromagnet is defined by the Hamiltonian
 
\begin{equation}
H_{\rm spin}= \kappa~\sum_{<x,y>}\sum_{A=1}^{N_F^2-1}J^A(x)J^A(y)
\label{afm}
\end{equation}
with $<x,y>$ nearest neighbor sites $x$ and $y$ on a square lattice
and the ``spin'' operators $J^A(x)$ in an irreducible representation
of the $SU(N_F)$ Lie algebra
\begin{equation}
\left[ J^A(x), J^B(y)\right]=if^{ABC}\delta(x,y)J^C(x)
\end{equation}
When the representation at each site is a rectangular Young Tableau
with $m$ rows and $N_C$ columns, it is convenient to represent the
spin operators by the fermion bilinears
\begin{equation}
J^A(x)=\sum_{\alpha=1}^{N_C}\sum_{a,b=1}^{N_F}
\psi^{\dagger}_{a\alpha}(x) T^A_{ab}\psi_{b\alpha}(x) 
\end{equation}
The fermions have the anticommutator,
\begin{equation}
\{ \psi_{a\alpha}(x),\psi^{\dagger}_{b\beta}(y)\}=
\delta_{ab}\delta_{\alpha\beta}\delta(x,y)
\label{fa}
\end{equation}
Constraints which project out the irreducible representation of the
spin algebra are
\begin{equation}
{\cal G}_{\alpha\beta}(x)\equiv
\sum_{a=1}^{N_F}
 \psi^{\dagger}_{a\alpha}(x)\psi_{a\beta}(x)-\delta_{\alpha\beta}
N_F/2~\sim~0~~\forall x
\label{gl1}
\end{equation}
${\cal G}_{\alpha\beta}(x)$ obeys the $U(N_C)$ Lie algebra, commutes
with the Hamiltonian and acts as the generator of gauge
transformations with gauge group $U(N_C)$.

The critical behavior of the antiferromagnet was examined by Read and
Sachdev \cite{read} using semiclassical methods.  The only free
parameters are the integers $N_C$ and $N_F$.  $N_C \ll N_F$ is the
classical limit of large representations, where the classical N\'eel
ground state is stable with the staggered spin order parameter
\begin{equation}
\mu_{ab}= (-1)^{\sum_i x_i}< \sum_{\alpha=1}^{N_C}
\psi_{a\alpha}(x)\psi_{b\alpha}(x)>
\label{op}
\end{equation}  
On the other hand, the limit $N_F \ll N_C$ is the quantum limit where
fluctuations are important and the system is in a spin disordered
state.  For both $N_C$ and $N_F$ large, they find a line of second
order phase transitions in the $(N_C,N_F)$ plane at $N_F={\rm
const.}\cdot N_C$ where the constant is a number of order one.

The relationship between the antiferromagnet and QCD is a very close
one.  There is an argument in ref. \cite{langmann} which maps the strong
coupling limit of lattice QCD onto the quantum antiferromagnet.  The lattice 
regularization of
the QCD Hamiltonian uses staggered fermions~\cite{kogut}. Since the
order parameters are identical, the N\'eel ordered states of the
antiferromagnet correspond to chiral symmetry breaking states of QCD.
Thus, the infinite coupling limit of QCD is identical to the quantum 
antiferromagnet.  A main difference between QCD with finite coupling
and the quantum antiferromagnet is that QCD allows a fermion kinetic energy term 
while retaining gauge invariance, whereas in the antiferromagnet, the fermions 
are not
allowed to move.  One could, however, regard the corrections to the strong
coupling limit of QCD as the addition of degrees of freedom and gauge
invariant perturbations in the quantum antiferromagnet which allow fermion
propagation.  In fact, it has even been suggested \cite{beran} that the 
additional degrees of freedom are generated dynamically.  

A common feature of both theories is that, aside from
$N_F$ and $N_C$ they have no free parameters. One could imagine adding
operators of the sort that, if their coupling constants is varied, can induce 
the chiral phase transition.  It is tempting to speculate that
these transitions fall into a universality class which can take into
account all such modifications, as long as they respect the symmetries
of the theory. Restricting attention only to those operators which
lead to a Lorentz invariant continuum limit, it was recently argued 
\cite{hamidian2} that the
universality class is described by the $4-\epsilon$ dimensional
Euclidean field theory,
\bea
S&=&\int d^{4-\epsilon}x\Bigg[ \frac{1}{2}{\rm
tr}(\nabla\phi) \cdot (\nabla\phi)\nonumber\\&&
+ {8\pi^2\mu^{\epsilon}\over
4!}\Big(\frac{g_1}{N_F^2}({\rm tr}\phi^2)^2+\frac{g_2}{N_F}{\rm
tr}\phi^4 \Big)
+{1\over4}{\rm tr}F_{\mu\nu}^2\nonumber\\&& 
+\bar\psi \Big( \gamma \cdot \nabla+i\mu^{\epsilon/2}e_1 \gamma \cdot A
+i\mu^{\epsilon/2}e_2 \gamma \cdot{\rm tr}A \nonumber\\&&
+\frac{\pi\mu^{\epsilon/2}y}{\sqrt{N_FN_C}}\phi\Big)\psi
\Bigg)
\label{action}
\eea
Here the scalar field $\phi$ is an $N_F\times N_F$ traceless Hermitean
matrix.  The 4-component spinor $\psi$ is an $N_F\times N_C$ complex
matrix and $A_\mu$ is a $U(N_C)$ gauge field.  In four dimensions this
model has Euclidean Lorentz invariance, C,P and T, discrete chiral
symmetry, ($\psi\rightarrow\gamma^5\psi$, $\phi
\rightarrow -\phi$) and global $SU(N_F)$ flavor.  
(\ref{action}) includes all operators which are marginal when $D=4$.

The evidence that (\ref{action}) describes the universality class
comes from previous work ~\cite{hamidian1} where a similar model where gauge
couplings are absent was studied. In Ref.~\cite{hamidian2} it was shown that the 
anomalous dimensions of operators computed in the model (\ref{action}) with
$e_i=0$ were identical to leading order in $1/N_C$ and $\epsilon$ to
those of a $2<D<4$ dimensional four-fermi theory.  That a $4-\epsilon$
dimensional Yukawa-Higgs theory has the same universal critical
behavior as lower dimensional four-fermi theories with the same
symmetries was originally suggested by Wilson ~\cite{wilson2}.  For the
case $N_F=1$, where the chiral symmetry is discrete, higher order
computations have been carried out ~\cite{zinn,rosenstein}. These results, as
well as those of lattice simulations, support the universality
hypothesis ~\cite{zinn}. It was argued in \cite{hamidian2} that (\ref{action}) 
represents the universality class of lower-dimensional four-fermi theories with 
$U(N_C)$ gauge invariance such that for
a large range of values of $(N_C,N_F)$, the chiral phase transition is
a fluctuation induced first order transition and when it is second
order, critical exponents are in principle computable in the $\epsilon$ 
expansion. 

To include the effects of QD amounts to adding new operators which can induce 
the chiral phase transition by varying the width, $\Delta$, of the 
impurity-probability distribution. This is achieved by adding 
\bea
\int d^d x (r + \delta r(x)) {\rm tr} \phi^2 (x)
\label{imaction}
\eea
to the aciton (\ref{action}). Here $\delta r(x)$ is the impurity field and $r$ 
is the mass of the field $\phi(x)$ (renormalized to zero at the critical point). 
The quenched partition function can be written by using the replica trick 
\cite{edwards}. Defining the impurity-probability distribution $P[\delta r]$ 
such that the replica average
$\langle\langle\delta r(x)\rangle\rangle = 0$ and
\bea
\langle\langle\delta r(x) \delta r(x')\rangle\rangle = \Delta 
\delta^{d-\epsilon_d} (x-x'),
\label{impcorr}
\eea
a cumulant expansion yields the (replica) action for the gauged Yukawa matrix 
field theory with QD
\bea
S_{\rm QD} &=& \sum_\alpha S_{\alpha}-\frac{\Delta}{2}\int d^dx \int d^d x' 
\delta^{d-\epsilon_d} (x-x') \nonumber\\&&\sum_\alpha {\rm tr}\phi_{\alpha}^2 
(x) \sum_\beta {\rm tr}\phi_{\beta}^2 (x').
\label{dopedaction}
\eea
Here $\alpha,\beta=1,\ldots,n$ are the replica indices and $S_\alpha$ has the 
same form as $S$ in (\ref{action}) with a replica index $\alpha$ for each field 
operator and $d=4-\epsilon$. In (\ref{dopedaction}) the impurities are 
correlated along $\epsilon_d$ dimensions, whereas in the remaining 
$d-\epsilon_d$ directions the impurity-correlation function (\ref{impcorr}) is 
zero for impurities at different points. It is important to note that the system 
is characterized by two dimensionalities, namely $\epsilon$ and $\epsilon_d$, 
which means that the perturbative expansion will, in general, involve a double 
$\epsilon$ expansion. Such a double $\epsilon$ expansion is necessary for a 
controlled perturbative description of the critical behavior (see Dorogovtsev in 
Refs. \cite{dorogovtsev} and \cite{boyanovsky}).

To investigate the critical behavior we first compute the $\beta$ functions for 
the model defined by (\ref{dopedaction}). To leading order, our calculations 
yield
\bea
\beta_1 &=& -\epsilon g_1 + \frac{N_F^2 + 7}{6 N_F^2} g_1^2 + \frac{2 N_F^2 - 
3}{3 N_F^2} g_1 g_2+\frac{N_F^2 + 3}{2N_F^2} g_2^2 + \frac{1}{2N_F} y^2 g_1 - 24 
g_1 \Delta,\\
\beta_2 &=& -\epsilon g_2 + \frac{2}{N_F^2} g_1 g_2 + \frac{N_F^2 - 9}{3 N_F^2} 
g_2^2 -\frac{3}{8 N_C N_F} y^4
+\frac{1}{2N_F} y^2 g_2,\\
\beta_y &=& -\frac{\epsilon}{2} y - \frac{3}{16 \pi^2} \frac{N_C^2-1}{N_C}e_1^2 
y - \frac{3}{8 \pi^2}e_2^2 y
+\frac{N_F^2 + 2 N_C N_F - 3}{16 N_C N_F^2} y^3,\\
\beta_{e_1} &=& -\frac{\epsilon}{2} e_1 - \frac{11N_C - 2N_F}{48 \pi^2} e_1^3,\\
\beta_{e_2} &=& -\frac{\epsilon}{2} e_2 + \frac{N_C N_F}{12 \pi^2} e_2^3,\\
\beta_\Delta &=& -{\tilde\epsilon}\Delta - 16 \Delta^2 + \frac{N_F^2 + 
1}{3N_F^2}g_1 \Delta + \frac{2N_F^2 - 3}{3 N_F^2} g_2 \Delta,
\eea
where ${\tilde\epsilon}=\epsilon + \epsilon_d$ and a factor of $2\pi^{\tilde 
d}\Gamma({\tilde d}/2)$, with ${\tilde d}=4-{\tilde\epsilon}$, has been absorbed 
into the definition of $\Delta$.
Fixed points occur at the zeros of the $\beta$ functions, $\beta(g_i^*)=0$. The 
fixed points are IR stable if all the eigenvalues of the stability matrix, 
$\partial \beta_i/\partial g_j |_{g=g^*}$, are positive. Second order phase 
transitions are possible when the RG trajectories flow to the IR stable fixed 
point in the region allowed by general stability conditions. Otherwise, the only 
allowed phase transition is a fluctuation induced first order one. Yamagishi 
\cite{yamagishi} formulated a criterion for this behavior. He showed that this 
occurs when the RG trajectory crosses the surface in coupling-constant space 
given by
\bea
{\cal P}_i(g,y,e) = 0,~~i=1,2,
\eea
where ${\cal P}_1(g,y,e)=(4-\epsilon)(g_1+g_2)+\beta_1+\beta_2$ and ${\cal P}_2 
= (4-\epsilon)(g_1 + (N_F/2)g_2) + \beta_1 + (N_F/2)\beta_2$ depending on 
whether $g_2>0$ or $g_2<0$, respectively. When the RG trajectories cross these 
surfaces two further conditions must be met. To ensure that the extremum is a 
local minimum, rather than a maximum, it is necessary that $D_i>0$ for $i=1,2$ 
where $D_1=(4-\epsilon)(\beta_1 + \beta_2) + \Sigma_i \beta_i \partial/\partial 
g_i (\beta_1 + \beta_2)$ or $D_2=(4-\epsilon)(\beta_1 + (N_F/2)\beta_2) + 
\Sigma_i \beta_i \partial/\partial g_i (\beta_1 + (N_F/2)\beta_2)$ depending on 
whether $g_2>0$ or $g_2<0$. 

For this minimum to have lower free energy than the one at the trivial $\phi=0$, 
it is necessary that the couplings at that scale obey
\bea
g_1+g_2<0~~{\rm or}~~g_1+N_F g_2/2<0.
\label{stabilityline}
\eea  
 
By using Eqns. (11)-(16) we have numerically solved for the RG trajectories in 
the six-dimensional space of couplings. In Fig.~1 we have plotted the projection 
of some of the RG trajectories onto the $g_1-g_2$ plane for the case $N_C=2$, 
$N_F=13$ and $\epsilon=\epsilon_d=0.1$. As in  the pure case, some of the RG 
trajectories satisfy the Yamagishi conditions for fluctuation-induced first 
order behavior before reaching the IR stable fixed point. Two examples of this 
are the middle dashed and lower solid curves corresponding to cases with and 
without QD respectively . 

To guarantee the consistency of the above analysis based on the WF $\epsilon$ 
expansion one needs to check for the Harris criterion \cite{harris} which 
requires the calculation of the critical exponent $\nu$ given by
\bea
\nu = \frac{1}{2-\gamma_2(g^*)}
\eea
where $\gamma_2(g^*)$ is the (mass) anomalous dimension. We have calculated 
$\gamma_2(g)$ in the presence of QD: 
\bea
\gamma_2 = \frac{N_F^2 + 1}{6 N_F^2} g_1 + \frac{2N_F^2 - 3}{6N_F^2} g_2 + 
\frac{1}{4N_F}y^2 - 4 \Delta
\eea
and checked for the conditition $\tilde{d} \nu \geq 2$ (the Harris criterion) at 
the fixed point $(g_1^*,g_2^*)$. Although we find that the Harris criterion is 
satisfied in the entire region $0 \leq \epsilon,\tilde{\epsilon} \leq 1$, an 
examination of the RG trajectories (see Fig.1) reveals that as in the pure case, 
except for a narrow range of parameters in the $(N_C,N_F)$ plane, the phase 
transition is fluctuation induced {\em first order}. The occurence of this 
phenomenon depends on the parameters $N_C$ and $N_F$ (see Ref. \cite{hamidian2} 
for details on a similar system without impurities) and typically there exists a 
critical impurity concentration, $\Delta_{\rm crit.}$, above which the phase 
transition will be fluctuation induced first order. In the range of parameters 
where the phase transition is second order, the gauged Yukawa matrix theory with 
(weak) quenched  disorder belongs to the universality class of the pure system. 
Since the specific heat exponent remains neagative, this is expected and further 
supports the consistency of this work with previous results. The physical 
quantum spin $j$ antiferromagnet corresponds to the parameter values $N_F=2$, 
$N_C=2j$. With $N_C=1(j=1/2)$ the non-Abelian field disappears and the resulting 
theory has no IR fixed point, indicating a first order phase transition. The 
case $N_F=2$, $N_C \geq 2$ (i.e., $j \geq 1$) belongs to the asymptotically free 
regime and hence these antiferromagnets cannot be analyzed by our techniques. In 
this domain in the $N_C$,$N_F$ parameter space we speculate that confinement is 
associated with a nonperturbative IR fixed point of the gauge coupling. In this 
case these antiferromagnets are likely to exhibit second order phase transition. 

In conclusion, we have shown that long wavelength quantum fluctuations can drive 
first order phase transitions in the presence of weak quenched disorder when 
naively none is expected. 
It is important to note that we have studied the effects of QD on the critical 
behavior of the {\em pure system} by first analyzing the relevance or 
irrelevance of disorder at the critical fixed point of the pure system and then 
carrying out a thorough examination of the stability of the pure system against 
quantum fluctuations. On the other hand, one may also study the crossover from 
the pure fixed point to the disorder fixed point. This is an important question 
whose examination, however, is beyond the scope of the present paper. In 
general, detailed calculations \cite{weinrib} show that when the pure fixed 
point is unstable (i.e. $\nu d<2$) the disorder fixed point is a critical fixed 
point with new critical exponents [In such instances the stability matrix 
typically has complex eigenvalues (see, e.g., Dorogovtsev in 
Ref.\cite{dorogovtsev}), which imply that the RG trajectories spiral into the 
fixed point basins of attraction, and give oscillatory corrections to scaling].
The results of this paper may help us better understand the nature of phase 
transitions and the underlying dynamics in low-dimensional systems with weak 
quenched disorder, such as those occuring in planar high-$T_C$ superconductors.

It is a pleasure to thank Michael~Ma for invaluable discussions and 
correspondence which culminated in this work. I would also like to thank 
T.H.~Hansson, E.~Langmann and L.C.R.~Wijewardhana for thoughtful comments and 
numerous discussions while this work was being completed.

This work is supported in part by the United States Department of Energy under 
Grant No. DE-FG02-84ER40153 and by the Swedish Natural Science Research Council.

\bibliographystyle{plain}

\noindent
Figure caption: Fig. 1: Projection of the RG trajectories in the six-dimensional 
coupling constant space onto the $g_1-g_2$ plane for $N_C=2$, $N_F=13$ and 
$\epsilon=\epsilon_d=0.1$. The dashed and solid curves correspond to theories 
with and without quenched disorder, respectively. The straight dashed line is 
defined by Eq. (18) and the lowest curve is a runaway trajectory.

\newpage
\thispagestyle{empty}
\begin{figure}[hbtp]
\centerline{\epsffile{2Ddpd.EPSF}}
\end{figure}

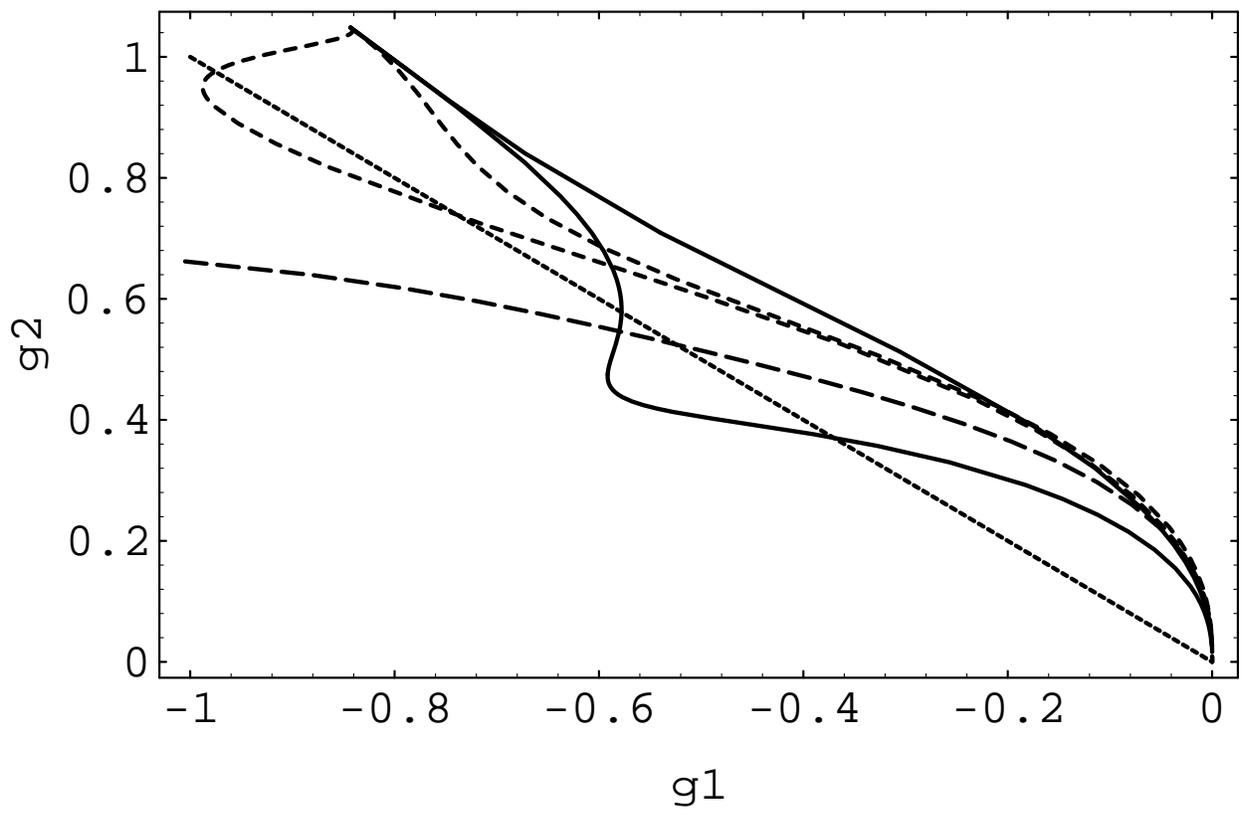
\end{document}